**Atomic-scale Modulation of Synthetic Magnetic Order in Oxide Superlattices**

*Seung Gyo Jeong, Sehwan Song, Sungkyun Park, Valeria Lauter, and Woo Seok Choi\**

S. G. Jeong, W. S. Choi

Department of Physics, Sungkyunkwan University, Suwon 16419, Korea

E-mail: choiws@skku.edu

S. Song, S. Park

Department of Physics, Pusan National University, Busan 46241, Korea

V. Lauter

Neutron Scattering Division, Neutron Sciences Directorate, Oak Ridge National Laboratory, Oak Ridge 37831, USA

Keywords: atomic-scale modulation, synthetic magnetic order, tunable magnetic noncollinearity, magnetic oxide superlattices, polarized neutron reflectometry

Atomic-scale precision control of magnetic interactions facilitates a synthetic spin order useful for spintronics, including advanced memory and quantum logic devices. Conventional modulation of synthetic spin order has been limited to metallic heterostructures that exploit RKKY interaction through a nonmagnetic metallic spacer; however, they face problems arising from Joule heating and/or electric breakdown. The practical realization and observation of a synthetic spin order across a nonmagnetic insulating spacer would lead to the development of spin-related devices with a completely different concept. Herein, we report the atomic-scale modulation of the synthetic spiral spin order in oxide superlattices composed of ferromagnetic metal and nonmagnetic insulator layers. The atomically controlled superlattice exhibit an oscillatory magnetic behavior, representing the existence of a spiral spin structure. Depth-sensitive polarized neutron reflectometry evidences modulated spiral spin structures as a function of the nonmagnetic insulator layer thickness. Atomic-scale customization of the spin state could lead the field one step further to actual spintronic applications.

# 1. Introduction

Synthetic spin order in magnetic heterostructures promotes novel spintronic functionalities[1] including colossal magnetoresistance,[2, 3] tunneling magnetoresistance,[4] topological Hall effect,[5] spin Hall effect,[6, 7] spin-wave propagation,[8] and terahertz spin-transfer torque.[9] In typical ferromagnetic (FM)/*nonmagnetic-metal* (*NM-M*)/FM heterostructures, the relative spin orientation between two FM layers can be modulated by the thickness of the NM-M layer, thereby realizing a synthetic magnetic order, which is useful for designing magnetic storage and logic devices.[1] This is generally understood by the Ruderman-Kittel-Kasuya-Yosida (RKKY) interlayer exchange interaction between the FM layers, which is mediated by the conduction electrons in the NM-M layer. In this scheme, the interaction strength oscillates as a function of the NM-M layer thickness.[1] In contrast, FM/*nonmagnetic-insulator* (*NM-I*)/FM heterostructures foster synthetic spin order, with unfamiliar exchange mechanisms other than the RKKY interaction.[10, 11] Recently, chiral phonon has been suggested to carry the spin information across the NM-I layer via strong spin-phonon and spin-orbit coupling,[11] similarly acting as the conduction electron across the NM-M layer for the RKKY interaction.[10, 11] It has been shown that the chiral phonons could mediate the interlayer exchange interaction leading to an oscillatory magnetization as a function the NM-I layer thickness, evidenced by confocal Raman spectroscopy.[10] If the synthetic spin order is atomically controllable in NM-I based heterostructures, the inherent limitations such as Joule heating and electric breakdown in NM-M based heterostructures can be resolved.

Polarized neutron reflectometry (PNR) is a favorable technique for investigating the atomic layer-resolved distribution of spins, particularly when combined with atomic-scale epitaxy of synthetic magnetic heterostructures.[12-15] Frist, the in-plane spin orientation in the magnetic layers can be obtained by comparing the non-spin-flip and spin-flip components of neutrons in the PNR spectra. Second, the depth profile of the spin orientation can be obtained by assessing the out-of-plane component of the wavevector transfer ($Q_z$) in the specular PNR. Third, application of the PNR to magnetic superlattices is further advantageous because the superlattice Bragg peak results in an enhanced reflectivity signal for the analyses.

In this study, we present an atomic-scale thickness control of the synthetic spin structures in oxide superlattices. We epitaxially deposited SrRuO$_3$ (SRO, FM layer)/SrTiO$_3$ (STO, NM-I layer) superlattices on (001)-oriented single crystal STO substrates using pulsed laser epitaxy.

As schematically shown in **Figure 1**a,b, the spiral spin structure of six-unit-cell-thick (1 u.c. ~0.4 nm) FM SRO layers was modulated by the $y$-u.c.-thick NM-I STO layers with ten repetitions, within the [6|$y$] superlattice. Consequently, $y$-dependent oscillatory net magnetization was obtained consistently from both the magnetization and the PNR measurements. We note that the magnetic easy axis of SRO thin films usually point to the out-of-plane direction. However, the $y$-dependent oscillatory behavior is only observed for the in-plane magnetization measurements. Since PNR also identifies the in-plane magnetization of the thin films, it is an ideal tool to characterize the important magnetic features of the superlattices in microscopic scale.

## 2. Results and Discussion

Figure 1c,d show X-ray reflectivity (XRR) and diffraction (XRD) results, respectively, validating the atomically defined periodicities of the SRO/STO superlattices. The XRR curves exhibit distinct Bragg peaks ($SL^{+n}$) and Kiessig fringes corresponding to the total thickness of the superlattice thin film, thereby indicating a well-defined periodic supercell structure with atomically sharp interfaces (Figure 1c). Figure S1 shows the XRR fitting results of the [6|4], [6|6], and [6|8] superlattices. Small decay slopes of the XRR indicate that the surface roughness of the superlattices is < 1 u.c., which is consistent with the topographic images obtained by atomic force microscopy (Figure S2). XRD $\theta$-$2\theta$ scans show coherent diffraction peaks ($SL^{\pm n}$) of the superlattices on the (001)-oriented single crystal STO substrates (substrate diffraction peaks marked by asterisks) (Figure 1d). With increasing $y$, the separation between the superlattice peaks decreases systematically, which indicates the atomically controlled periodicities ($\Lambda_{SL}$). When thickness of the $x = 6$ u.c. of SRO layer is assumed to be fixed at 2.358 nm (obtained from epitaxially strained single SRO thin films on STO substrates), the estimated thicknesses of the $y$ u.c. of STO layer are 0.768, 1.573, 2.318, 3.078, and 6.957 nm for the [6|2], [6|4], [6|6], [6|8], and [6|18] superlattices, respectively. These values are obtained from the Bragg's law, $\Lambda_{SL} = 2\pi(Q^n - Q^{n-1})^{-1}$, where $n$ and $Q^n$ denote the superlattice peak order and the $Q_z$ position of the $n$th-order superlattice peak, respectively. The deviation between the target and measured thicknesses is smaller than half a u.c. (< 0.2 nm), thus manifesting structurally well-controlled superlattices. Finally, Figure 1e representatively shows the X-ray reciprocal space mapping of [6|8] superlattice around the (103) Bragg diffraction peak of the STO substrate, confirming a fully strained state.

The in-plane magnetization of the [6|y] superlattices shows an unexpected oscillation as a function of y at low-temperature (-T) and low-magnetic- (H-) fields. Field-cooled M (T) curves of the [6|y] superlattices were measured at 0.01 T of H-field along the in-plane direction (**Figure 2**a). In addition to a robust FM transition at approximately 130 K (consistent with 6 u.c. SRO single layer on the STO substrate[16, 17]), the in-plane M shows a peak as T is further reduced, indicating that the FM order is disturbed. The M (H) curves at low T exhibit a double hysteresis with a large coercive field ($H_c$) of approximately 1.8 T (see the inset of Figure 2a), which supports the disturbed magnetic order in the ground state. Moreover, the M (H) curve of y = 18 u.c. superlattice shows a typical single hysteresis loop with a small value of $H_c$, but with a saturation M ($M_s$) of ~0.3 $\mu_B$/Ru similar to those of y = 6 and 8 u.c. superlattices, which indicates the suppression of the interlayer exchange interaction at a sufficiently thick NM-I layer. We note that the enhancement of $M_s$ for y = 4 u.c. superlattice might originate from the structural modulation (orthorhombic to tetragonal) in the SRO layers with decreasing y.[16] The oscillation of the in-plane M at 5 K and 0.01 T is shown as a function of y in Figure 2f.[10] The magnetization results suggest the existence of an interlayer exchange interaction across the NM-I STO layer and the possible unconventional synthetic magnetic order in the SRO/STO superlattices. The interlayer exchange interaction strength is estimated as $J = t_{FM}H_cM_s$, where $t_{FM}$ is the thickness of the FM SRO layer. Figure S3 shows the y-dependent oscillatory behavior of J, which depends on both $H_c$ and $M_s$. This confirms the unconventional character of the interlayer exchange interaction between FM SRO layer across the NM-I STO layer, which is supposed to originate from the chiral phonon.[10,11] The J values in SRO/STO superlattices are approximately two orders of magnitude less than those for the RKKY interaction at the same thickness of the NM-M layer.[18] Although conventional analyses of the M (H) curve would be useful to examine the macroscopic strength of magnetic interaction, we note that the observed magnetic order is rather fragile and is easily destroyed even in a moderate H-field. Therefore, we will focus on the low H-field region of 0.01 T.

To understand the unexpected y-dependent oscillatory magnetization, we examine a simple model, wherein spins in SRO layers have a rotation angle $\phi$ with respect to the spins in the adjacent SRO layer. Let us first assume that each SRO layer has the same uniform in-plane $M_i$ value (i is the layer index). Although it has been reported that spin ordering may differ depending on the position away from the interface within magnetic heterostructures,[19,20] the modulation is generally small in atomically defined heterostructures with symmetric interfaces. Second, we assume that $\phi$ depends linearly on $y$, that is, with an increase in the NM-I layer thickness, the rotation of $M_i$ increases, as in the case of the RKKY with the NM-M layer. Figure 2b–d shows a schematic representation of the model. We estimate the sum of the projections of $M_i$ ($y$, $\phi$) along a certain in-plane direction of the $H$-field for each SRO layer to simulate the results of $y$-dependent $M$ because the magnetization measurement gives only the average scalar $M$ value of the entire superlattice along the $H$-field direction. The average $M$ value is determined by the relation $M = \sum_i M_i (\phi) = \sum_i [M_a + M_b \cos((i-1)\phi)]$, where $M_a$ and $M_b$ are constants. Next, we compare the result with that obtained from the experiment and calculate the sum of the squared errors (SSE) as a function of $\phi$, defined as [(Simulated $M$) – (Measured $M$)]$^2$ (Fig. 2(e)). In particular, for $y = 2$ u.c. (~0.8 nm), $\phi$ = ~80 and ~100° results in the lowest SSE values within the spiral spin models with $M_a = 0.070$ $\mu_B$/Ru and $M_b = 0.629$ $\mu_B$/Ru. Figure 2(f) shows that the spiral spin models with $\phi$ = ~200°, 300°, and 400° (40°) for the [6|4], [6|6], and [6|8] superlattices consistently describe the experimental $y$-dependent oscillatory behavior of $M$ at 5 K. Note that the underestimation of $M$ for the [6|4] superlattice might be due to the absence of the decaying term with increasing $y$ in our model, which would further complicate the model.

PNR is employed to clarify the complex synthetic spin structure suggested by the magnetization measurement discussed earlier (**Figure 3** and **Figure 4**). Figure 3a schematically shows that a collimated polychromatic neutron beam is incident on the film surface at a grazing angle ($\alpha$). The PNR signal was measured as a function of $Q_z = 4\pi\sin(\alpha)/\lambda$ along the out-of-plane direction ($z$-direction in Figure 3a), where $\lambda$ is the neutron wavelength. We assume that a homogeneous in-plane $M$ vector ($\vec{M}$) is rotated by angle $\phi$ within the $xy$-plane, defining the in-plane $M_x$ and $M_y$ components. Then the PNR signal would include both non-spin-flip ($R^{++}$ and $R^{--}$) and spin-flip ($R^{+-}$ and $R^{-+}$) contributions.[21] $R^{++}$ and $R^{--}$ are defined as $1/4|(r_+ + r_-) + (r_+ - r_-)\cos\phi|^2$ and $1/4|(r_+ + r_-) - (r_+ - r_-)\cos\phi|^2$, respectively, and $R^{+-}$ and $R^{-+}$ are described as $1/4|r_+ - r_-|^2 \sin^2\phi$, respectively, where $r_\pm$ are the complex reflection amplitudes for the up and down spins.[13] Because the simulated experimental signal with spin-flip polarization ($R^{+-}$ and $R^{-+}$) is rather small for our samples, and hence, required a long measuring time to obtain high statistics, we did not employ spin-flip polarization analyses. Therefore, the spin-flip reflection was taken into account only in the data analyses as follows. Our results show spin-up and spin-down polarizations of the PNR spectra, $R^+ = R^{++} + R^{+-}$ and $R^- = R^{--} + R^{-+}$, which describe depth-sensitive spin-vector rotation in the SRO/STO superlattices.

To minimize the number of parameters in PNR fitting, we confirm the prerequisite structural parameters and $M_s$ values of the SRO/STO superlattices. First, the structural parameters, including the interface roughness, thickness, and density of each layer, are obtained from non-polarized neutron reflectivity data at 300 K when the sample is nonmagnetic. The top panels of Figure 3c,e,g show the unpolarized neutron reflectivity and fitting results for the $y = 4, 6,$ and 8 superlattices, respectively, which are consistent with the XRR results shown in Figure 1c and Figure S1. The structural parameters of PNR analyses at 300 K have been confirmed with those of XRR fitting. We summarize the structural fitting parameters of the XRR and PNR results with the measured thickness consistently obtained by using XRR, PNR, and scanning transmission electron microscopy[10] in Table S1. These results manifest the atomically well-controlled SRO/STO superlattices with the minimized interdiffusion. Second, to confirm the $M_s$ values of the superlattices, we measured the PNR spectra at 85 K with a 1 T ($> H_c$ at 85 K) of in-plane $H$-field. The $M(H)$ curves at 85 K exhibit a single hysteresis loop as for a typical FM, as shown in Figure 3b. At this temperature with 1 T of $H$-field, the net $M$ is almost saturated (net $M \sim M_s$), which indicates that the in-plane rotation is relatively suppressed ($\phi \sim 0°$). The bottom panels of Figure 3c,e,g show the PNR spectra at 85 K in a 1 T field. The insets in Figure 3c,e,g highlight the differences in PNR intensity between $R^+$ and $R^-$ near the Bragg peaks of each superlattice. This clear separation is consistently observed in both the experimental result and fit, thereby indicating a robust FM order of the SRO layers at a high $H$-field at 85 K. Figure 3d,f,h show the nuclear and magnetic scattering length densities (SLD) fitted using GenX for [6|4], [6|6], and [6|8] superlattices, respectively.[22] They show an atomically well-defined periodic structure and ferromagnetically aligned $M_i$ vectors in each SRO layer at 85 K with a 1 T field. The obtained $M_s$ values are estimated to be 0.4 $\mu_B$/Ru for [6|4] superlattice and 0.3 $\mu_B$/Ru for [6|6] and [6|8] superlattices, respectively, which are highly consistent with the $M(H)$ curves in Figure 3b.

In the ground state (experiments at 5 K and 0.01 T of $H$-field), the PNR spectra reveal a modulated synthetic spiral spin structure suggested by magnetization measurements. Figure 4a shows the simulation result of a synthetic spiral spin structure with the best fit, which is consistent with the experimental PNR result. For a single magnetic film, the PNR spectrum at small $Q_z$ is influenced by the direct beam, whereas with increasing $Q_z$, the reflectivity decays exponentially such that the experimental data may be affected by the background signal.[12] In contrast, as briefly discussed previously, superlattice Bragg peaks provide a better signal-to-noise ratio even at finite $Q_z$ values due to the coherent periodic structure.[23] Therefore, we primarily compare the PNR data and simulation results near the $Q_z$ values associated with the superlattice Bragg peaks. Figure S4 summarizes PNR simulations of the SRO/STO superlattices for a typical FM model (left panels), a synthetic collinear antiferromagnetic (sAFM) model with $\phi = 180º$ (middle panels), and a synthetic spiral spin structure (right panels). The simulated spectra of collinear FM and sAFM models have two distinctions from the experimental PNR spectra: (1) The collinear sAFM structure leads to doubling of the magnetic unit cell of the superlattices. This doubling causes a significant discrepancy between $R^+$ and $R^-$ at half $Q_z$ of the superlattice Bragg peak.[24] As highlighted by the red rectangle in Figure S4a, this model produces spectra that are inconsistent with the experimental PNR spectra. (2) The collinear FM model results in a strong intensity contrast between the $R^+$ and $R^-$ in the superlattice Bragg peak due to the difference in the scattering of spin up and spin down neutrons of periodic superlattice structures.[24] Note the similarity to the PNR spectra at 85 K with robust FM ordering. As highlighted by the blue rectangle in Figure S4b, the experimental PNR spectra exhibit negligible separation in the superlattice Bragg peaks. Hence, the PNR experimental results show that a collinear spin configuration is unlikely in SRO/STO superlattices. Instead, the spiral spin structure model best describes the experimental PNR spectra among the considered simple models, confirming the in-plane spin rotation in the SRO/STO superlattice as a function of $y$.

## 3. Conclusion

In summary, we presented the modulation of synthetic magnetic order in FM/NM-I/FM superlattice via atomic-scale precision thickness control. To customize the in-plane rotation of spin vectors within ferromagnetic SRO layers, we atomically controlled the thickness $y$ of the NM-I (STO) layer within the SRO/STO superlattices. The oscillatory in-plane magnetic behavior as a function of $y$, determined by magnetization measurements, indicates the presence of the synthetic magnetic order in the superlattices. The PNR result manifests the depth profile of spin vectors within SRO layers with varying rotation angle $\phi$ and its controllability via the precise control of $y$. Both magnetization and PNR results consistently confirm that the atomically controlled thickness of the NM-I layers effectively changes the rotation angle of the spiral spin structures in the FM layers within the FM/NM-I superlattices. Our approach suggests an atomic-scale control knob for modulating synthetic spin spiral structures for future spintronics.

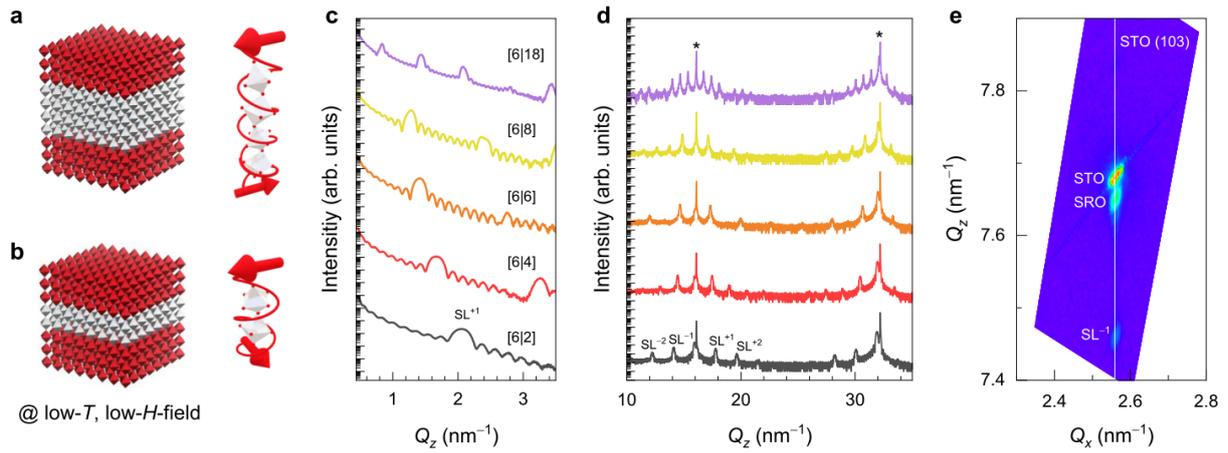

**Figure 1.** Atomically controlled SRO/STO superlattices for controlling synthetic magnetic order. a) and b) Schematic illustrations of modulation of synthetic magnetic order in FM (red)/NM-I (gray)/FM (red) heterostructures via atomic-scale precision thickness control of NM-I layer. c) X-ray reflectivity and d) $\theta$-$2\theta$ scan results of the atomically well-defined superlattices. e) X-ray reciprocal space mapping of [6|8] superlattice representatively shows the fully strained state.

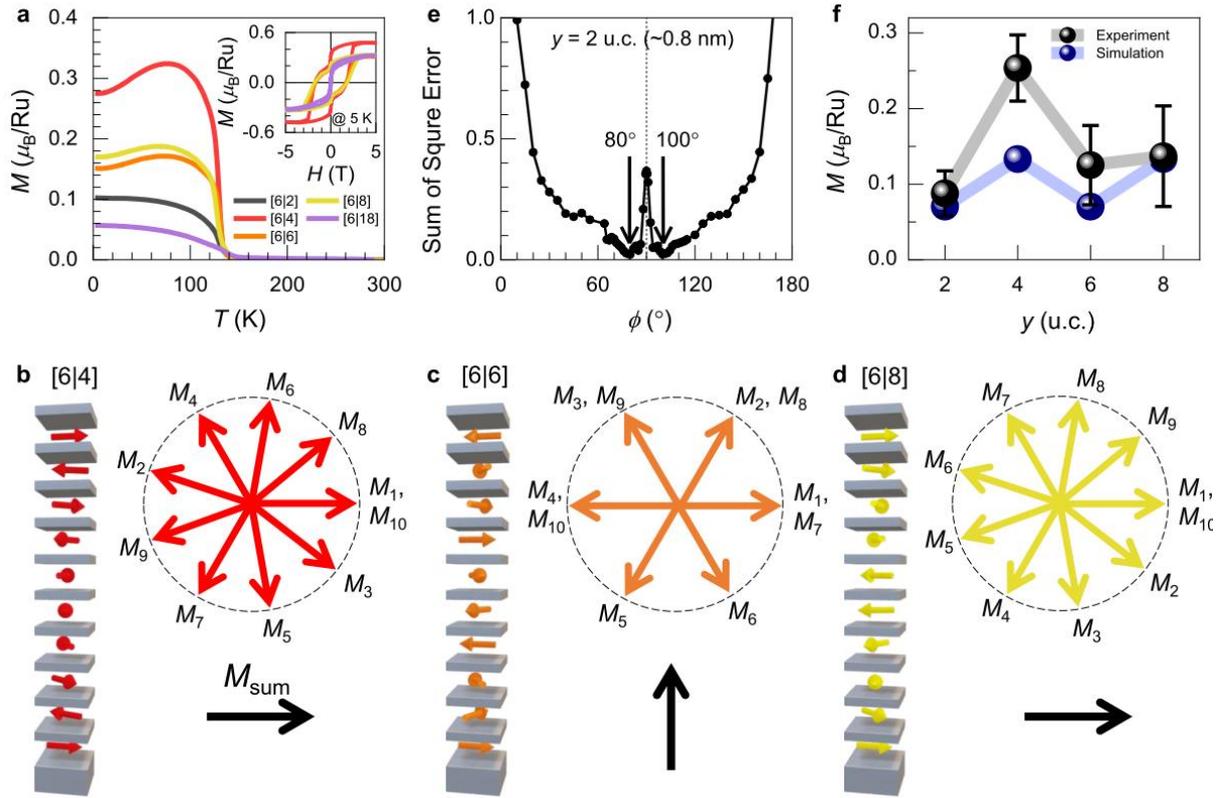

**Figure 2.** Synthetic in-plane magnetic behavior of SRO/STO superlattices. a) Field-cooled $M$ ($T$) curves of superlattices with 0.01 T of in-plane $H$-field. The inset shows the $M$ ($H$) curves of superlattices at 5 K. Schematic representation of in-plane $M_i$ vector of the SRO layer for b) [6|4], c) [6|6], and d) [6|8] superlattices. External $H$-field applied along the $y$-direction. The red arrows in the bottom panels denote the summation of $M_i$ vectors ($M_{sum}$). e) Sum of square error values as a function of $\phi$ for $y$ = 2 u.c. superlattice, determined by [(Simulated $M$) – (Measured $M$)]$^2$. f) Summary of experimentally measured $M$ and the simulated $M$ values as a function of $y$. The error bars indicate the experimental deviation from two different datasets.

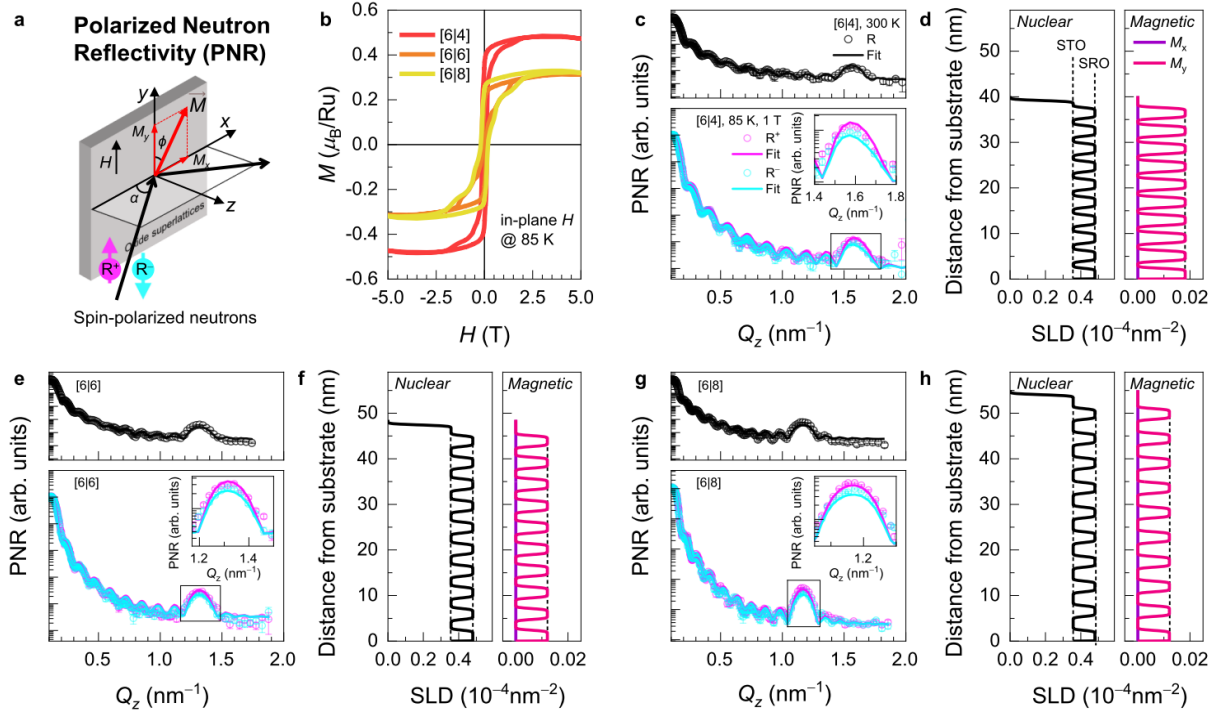

**Figure 3.** PNR results of the FM state of SRO/STO superlattices at 85 K and 1 T of *H*-field. a) Schematic representation of a PNR configuration with oxide thin films. b) *M* (*H*) curves of [6|*y*] superlattices with different *y* at 85 K. PNR spectra at 300 K and 85 K with 1 T *H*-field for c) [6|4], e) [6|6], and g) [6|8] superlattices. The insets show the extended PNR spectra near the Bragg peaks of each superlattice. Nuclear and magnetic SLD of d) [6|4], f) [6|6], and h) [6|8] superlattices. The vertical dashed lines in SLD are eye guides.

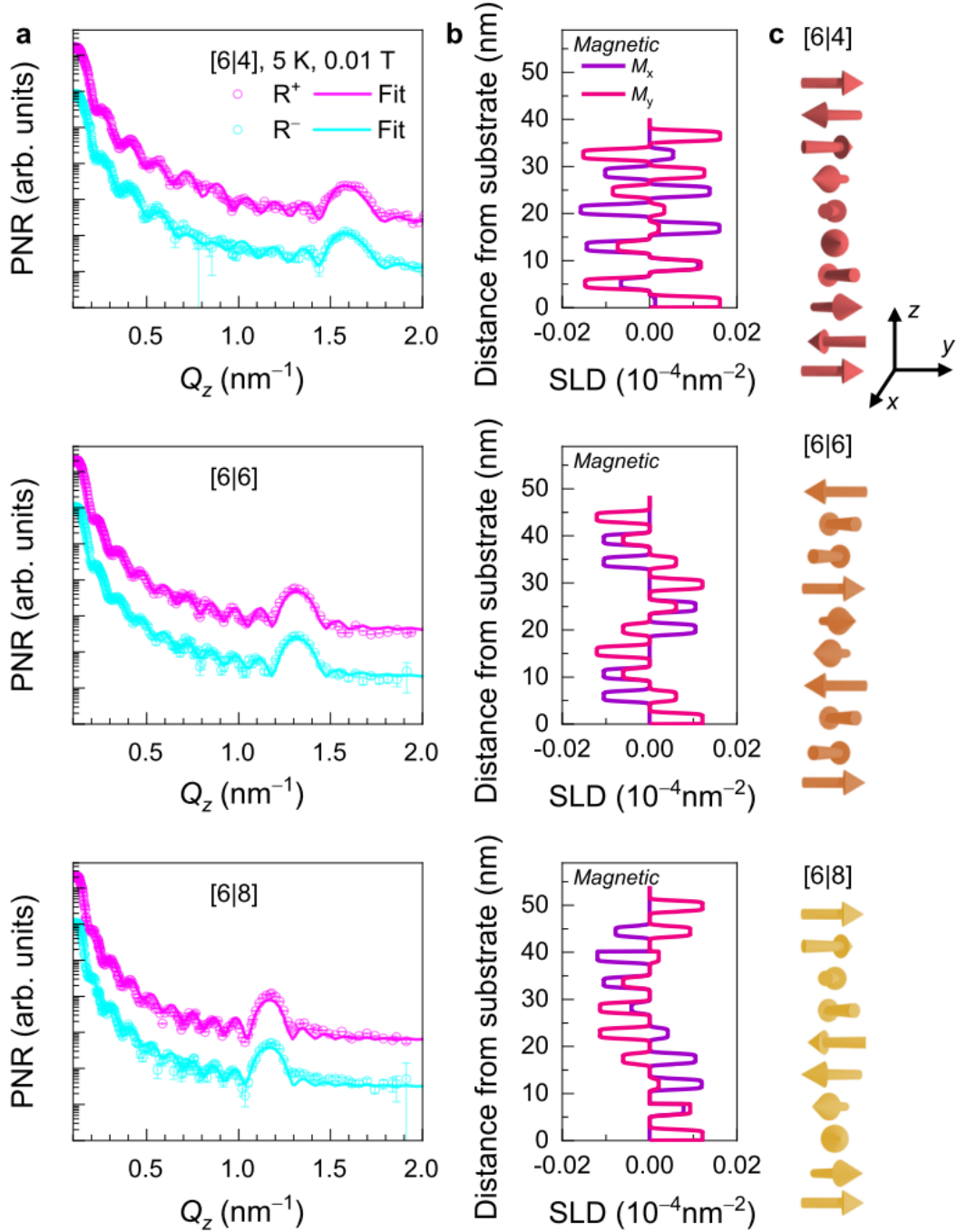

**Figure 4.** *y*-dependent in-plane rotation of synthetic spiral spin structures in SRO/STO superlattices at 5 K and 0.01 T of *H*-field. a) PNR spectra at 5 K and 0.01 T of *H*-field for [6|4], [6|6], and [6|8] superlattices. b) Magnetic SLD and c) schematic representation of the synthetic spin order of [6|4], [6|6], and [6|8] superlattices. The result evidences that the atomic-scale modulation of the synthetic spiral spin structure in SRO/STO superlattices as a function of *y* consistent with magnetization measurements.

## 4. Methods

*Atomic-scale epitaxy*: To validate the atomically designed FM/NM-I/FM heterostructures, we deposited epitaxial SRO/STO superlattices on (001)-oriented single crystal STO substrates by using pulsed laser epitaxy.[10, 17, 25–28] We fixed the number of laser pulses to grow the SRO layers and systematically changed the number of pulses to modulate $y$. We controlled the six u.c. of the SRO layers and $y$ u.c. of the STO layers with ten repetitions, i.e., the [6|$y$] superlattice. Before deposition, we treated the surfaces of the STO substrates using buffered HF and annealed them at 1000 °C for 6 h under atmospheric conditions. We used stoichiometric ceramic SRO and STO targets and an excimer (KrF) laser (248 nm; IPEX 868, LightMachinery) with a 1.5 J/cm$^2$ of fluence and a repetition rate of 5 Hz. The temperature and oxygen partial pressure employed for the stoichiometric growth of both SRO and STO layers were 750 °C and 100 mTorr, respectively. To characterize the atomically controlled periodicity of superlattice, we performed XRR and XRD $\theta$-$2\theta$ measurements using a high-resolution PANalytical X'Pert and Rigaku SmartLab X-ray diffractometer with Cu K-$\alpha_1$.

*Magnetization measurement*: The field-cooled $M$ ($T$) curves of the SRO/STO superlattice along the in-plane direction were measured using a magnetic property measurement system (MPMS, Quantum Design). The in-plane $M$ ($H$) curves of the SRO/STO superlattice were measured at 5 and 85 K. We did not observe any exchange bias effect in $M$ ($H$) curves of the SRO/STO superlattice. When we estimated the orientation of the spin vectors to describe the oscillatory $M$ behavior, we included the layer-repetition-dependent $M$ of superlattices as well (Figure S5).[10]

*Polarized neutron reflectometry*: The PNR experiments were performed on the time-of-flight Magnetism Reflectometer (BL-4A) at the Spallation Neutron Source at Oak Ridge National Laboratory (SNS, ORNL).[29] We used the closed-cycle refrigerator systems with a Bruker electromagnet to induce an in-plane $H$-field. We used polarized neutron beam with polarization efficiencies from 99 to 97.5% and $\lambda$ within a band of 2 to 8 Å. The PNR spectra with nuclear and spin structures were fitted using GenX.[22]

## Supporting Information

Supporting Information is available from the Wiley Online Library.

**Acknowledgements**

This research used resources at the Spallation Neutron Source, a Department of Energy (DOE) Office of Science User Facility operated by the Oak Ridge National Laboratory. We gratefully acknowledge Dr. Haile Ambaye (Spallation Neutron Source) for technical support with PNR experiments and data processing. We also thank Core Research Facilities, Pusan National University for MPMS. This research was supported by the Basic Science Research Programs through the National Research Foundation of Korea (NRF-2020K1A3A7A09077715, 2021R1A2C2011340, and 2022R1C1C2006723).

Atomic-scale precision epitaxy and microscopic observation let us customize the synthetic magnetic order useful for spintronic applications. However, conventional approaches have been limited to metallic heterostructures, which have Joule heating and/or electric breakdown. Here, we report atomic-scale modulation of synthetic magnetic order across the insulating spacer, observed by a polarized neutron reflectometer. This approach can yield novel controllability of magnetic order across insulating spacers.

Seung Gyo Jeong, Sehwan Song, Sungkyun Park, Valeria Lauter, and Woo Seok Choi*

Atomic-scale Modulation of Synthetic Magnetic Order in Oxide Superlattices

ToC figure

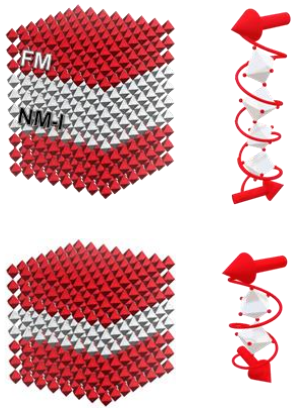

# Supporting Information

Atomic-scale Modulation of Synthetic Magnetic Order in Oxide Superlattices

*Seung Gyo Jeong, Sehwan Song, Sungkyun Park, Valeria Lauter, and Woo Seok Choi\**

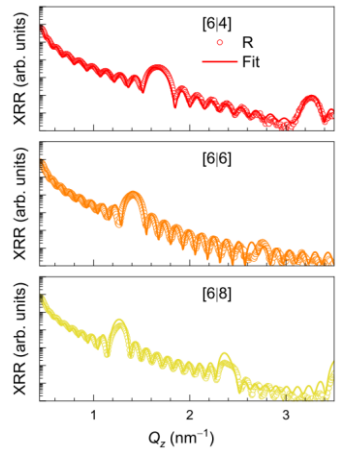

**Figure S1.** XRR fitting results of [6|$y$] superlattices with different $y$. The symbols (solid lines) are the experimental data (fit) of the XRR data.

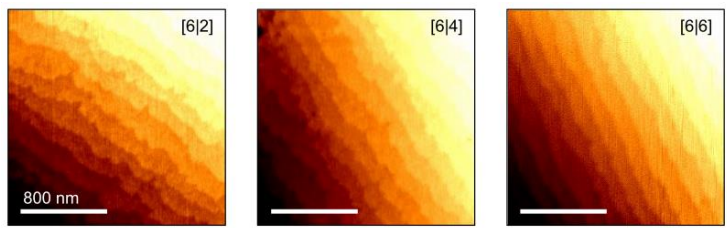

**Figure S2.** Atomic force microscopy images of [6|$y$] superlattices with different $y$ shows typical step and terrace structures indicating atomically flat surfaces of the samples.

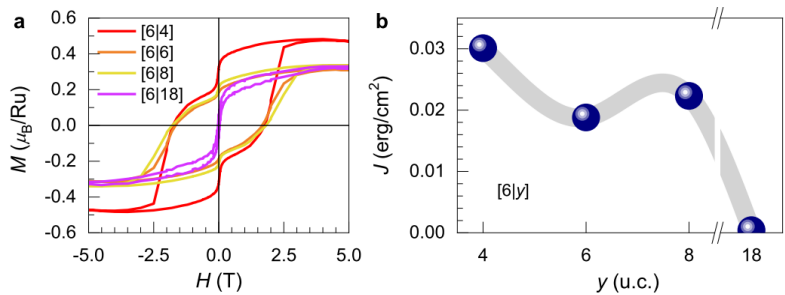

**Figure S3.** a) In-plane $M$ ($H$) curves of [6|$y$] superlattices with different $y$ at 5 K. b) Estimated $J$ of superlattices as a function of $y$.

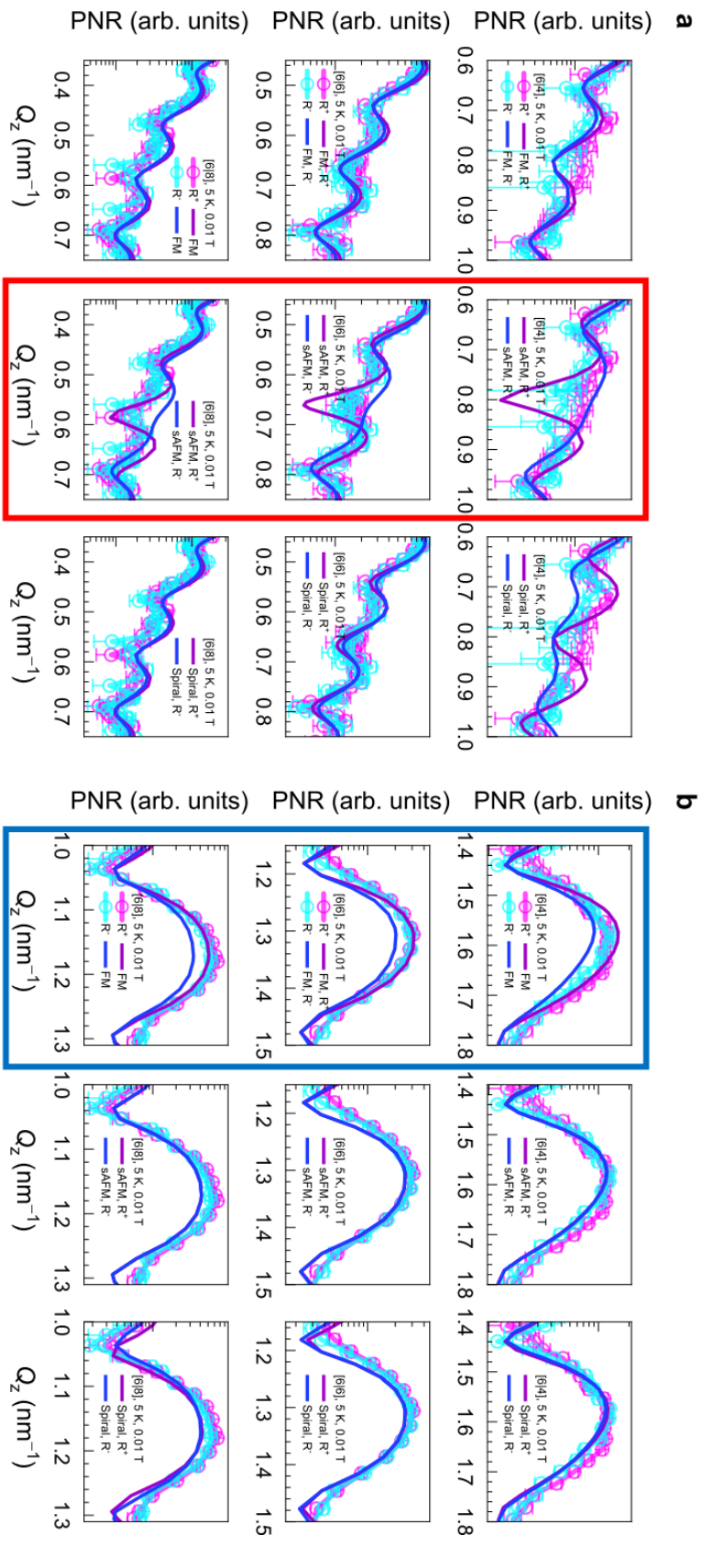

**Figure S4.** Extended PNR spectra with three different spin models (FM, sAFM, and spiral model, see Results and Discussion) at 5 K and 0.01 T of in-plane *H*-field near a) the half of the Bragg peaks and b) Bragg peaks of each superlattice. The red and blue rectangles highlight distinctions between collinear FM and sAFM models and PNR spectra results, respectively.

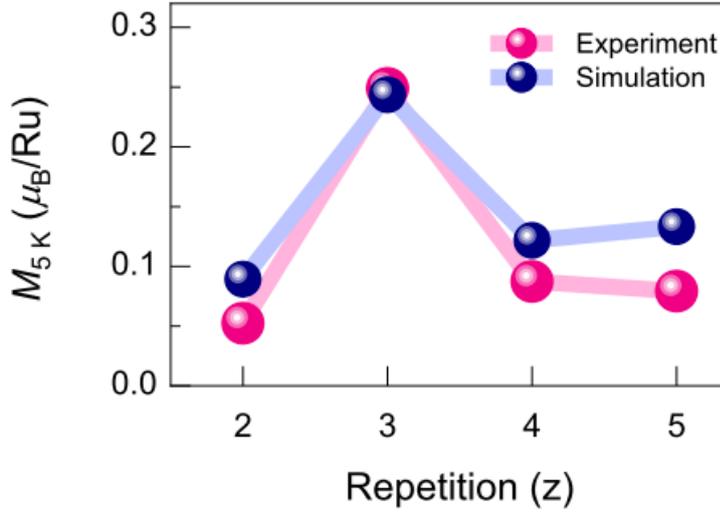

**Figure S5.** Comparison between the measured *M* and the simulated *M* values at 5 K for layer-repetition- (*z*-) dependent [6|4]$_z$ heterostructures.

**Table S1.** Summary of fitting parameters for XRR and PNR analyses and comparison measured thickness of SRO/STO superlattices

| Samples | XRR | | | | PNR | | |
|---|---|---|---|---|---|---|---|
| | Roughness (nm) | | Density (g/cm$^3$) | | Roughness (nm) | Density (f.u./Å$^3$) | |
| | SRO | STO | SRO | STO | | SRO | STO |
| [6\|4] | 0.14 | 0.23 | 5.9 | 4.81 | 0.2 | 0.0153 | 0.017 |
| [6\|6] | 0.38 | 0.19 | 6 | 4.12 | | | |

| Samples | Target thickness (nm) | Measured thickness (nm) | | |
|---|---|---|---|---|
| | | XRR | PNR | STEM |
| [6\|2] | 31.366 | 31.256 | - | 32.590 |
| [6\|4] | 39.176 | 39.306 | 39.134 | - |
| [6\|6] | 46.986 | 46.756 | 47.430 | - |
| [6\|8] | 54.796 | 54.356 | 53.930 | 54.930 |